# Stability, structural, elastic and electronic properties of RuN polymorphs from first-principles calculations

**V.V. Bannikov, I.R. Shein,\* A.L. Ivanovskii**

*Institute of Solid State Chemistry, Ural Branch, Russian Academy of Sciences, Pervomaiskaya St., 91, Ekaterinburg, 620990 Russia*

**Abstract**

First-principles calculations through a FLAPW-GGA method for six possible polymorphs of ruthenium mononitride RuN with various atomic coordination numbers CNs: cubic zinc blende (ZB) and cooperite PtS-like structures with CNs = 4; cubic rock-salt (RS), hexagonal WC-like and NiAs-like structures with CNs = 6 and cubic CsCl-like structure with CN = 8 indicate that the most stable is ZB structure, which is much more preferable for RuN than the recently reported RS structure for synthesized RuN samples. The elastic and electronic properties of ZB-RuN were investigated and discussed in comparison with those for RS-RuN polymorph.

\* Corresponding author. Tel.:  +7 343 3744453; fax: +7 343 3744495
E-mail address: shein@ihim.uran.ru (I.R. Shein)

**1. Introduction.**

Superhard and ultra-incompressible materials are of great practical significance and are useful in a variety of industrial applications such as abrasives, cutting tools, coatings etc. A promising approach to design such materials is to combine transition metals possessing a high valence electron density and high bulk modules, with small, covalent bond-forming sp atoms such as boron, carbon, nitrogen or oxygen [1].

Attempts to synthesize or theoretically predict new superhard and ultra-incompressible materials are the subject of intensive current researches, and great efforts are now focused on the synthesis and characterization of hard materials in the systems: platinum metals (PMs)-(B,C,N,O), see reviews [2,3].

Recently the ruthenium mononitride RuN was synthesized and declared as cubic (rock-salt-like, RS) phase [4]. However, it was found theoretically [5] that RS RuN is mechanically unstable. In addition, for this crystal the magnetic behavior was predicted [5]. The structural and electronic properties have been investigated also for RuN in the zinc-blende (ZB) structure [6] and it has been established that the cohesive energy for ZB phase is higher than for RS phase. Thus, the results [5,6] casts doubt on the proposed [4] rock-salt-like structure of ruthenium nitride. Moreover, as is known, the 3d (Fe) and 5d (Os) analogues of ruthenium are more inclined to form with nitrogen (or with carbon) the stable diamond-like polymorphs with four-fold atomic coordination than rock-salt-like structures with octahedral coordination [7-9].

Motivated by this background we have performed the first-principles calculations for comparative examination of relative stability, the lattice parameters, electronic and elastic properties for six possible polymorphs of ruthenium mononitride RuN. These six structures represent the most typical variants of atomic coordination (of atomic coordination numbers - CNs): cubic ZB and cooperite PtS-like structures with CNs = 4; cubic RS, hexagonal WC-like and NiAs-like structures with CNs = 6 and cubic CsCl-like structure with CN = 8.

## 2. Computational details.

The calculations were performed by means of full-potential LAPW method implemented in WIEN2k code [10]. The generalized gradient approximation (GGA) in the PBE form [11] was used for the exchange-correlation potential. The *muffin-tin* sphere radii were chosen to be 2.0 a.u. and 1.6 a.u. for Ru and N, respectively. The energy cutoff was taken as $-8.0$ Ry, and Ru($4s^2 4p^6 4d^7 5s^1$), N($2s^2 2p^3$) were treated as valence slates. The plane-wave expansion with $R_{MT} \times K_{MAX}$ was equal to 7, and $k$ sampling with 15×15×15 $k$-points mesh in the Brillouin zone was used. The Blöchl's modified tetrahedron method [12] was employed for the densities of states (DOS) calculations. We used the standard «volume-conserving» technique in the calculation of stress tensors on strains applied to the equilibrium structure to obtain the elastic constants $C_{ij}$ of RuN cubic modifications, see [13] for details.

## 3. Results and discussion.

As the first step we calculate the total energy ($E_{tot}$) *versus* the volume and find the ground state energies and the equilibrium lattice parameters for all six possible RuN polymorphs. These results (Fig. 1 and Table 1) allow us to make the following conclusions.

(i) All six RuN structures keep their initial symmetry after full geometry optimizations;

(ii) The relative stability of the examined RuN structures is: ZB-RuN > PtS-RuN > NiAs-RuN > WC-RuN > RS-RuN > CsCl-RuN, see Table 1; *i.e.* the stability of polymorphs grows with reduction of their CNs;

(iii) Four examined polymorphs (ZB-, PtS-, NiAs- and WC-like) are more stable than the declared cubic RS-RuN [4];

(iv) Moreover, among octahedral coordinated structures with CN=6 the cubic RS-RuN is less stable than hexagonal WC-like polymorph.

Let us note also that the spin-polarized calculations have shown that only RS-RuN became magnetic with considerable atomic magnetic moments (see also [13]), whereas all of the others RuN polymorphs are predicted to be non-magnetic. Besides, the ground state of magnetic RS-RuN is lower in energy only at about 0.02 eV/f.u. in comparison with non-magnetic state, and does not change the above stability sequence of considered polymorphs.

The lattice parameters for six RuN polymorphs are summarized in Table 1. We found that the calculated parameter $a^{calc}$ = 4.555 Å of the most stable ZB-RuN is in reasonable agreement with the experimental $a^{exp}$ = 4.445 Å [4]; with deviation $\Delta = (a^{calc} - a^{exp})/a^{exp} \times 100\%$ less than 2.5%, and some overestimation of the parameter $a^{calc}$ is due to the well known peculiarity of GGA calculations. Oppositely, the calculated parameter $a^{calc}$ = 4.313 Å for RS-RuN is underestimated in comparison with $a^{exp}$. These arguments also allow us to assert, that the most likely structure of synthesized RuN [4] may be the ZB-RuN structure.

For the further understanding the distinctions between ZB and RS polymorphs, their mechanical parameters such as the elastic constants $C_{11}, C_{12}$ and $C_{44}$, the bulk modules $B=(C_{11}+2C_{12})/3$, the shear modules $G=C_{44}$ and the tetragonal shear modules $G'=(C_{11}-C_{12})/2$ are evaluated, Table 2. It was found that $C_{ij}$ constants of ZB-RuN satisfy the generalized criteria of mechanical stability for cubic crystals ($C_{11}-C_{12}>0$, $C_{11}+2C_{12}>0$, $C_{44}>0$), whereas for RS-RuN (for which $C_{44}<0$) both G and G' are negative indicating its instability with respect to shear and tetragonal distortions. For ZB-RuN $B > G > G'$; this implies that the parameter limiting the mechanical stability of this material is the tetragonal shear modulus. According to Pugh's criteria [14] the material should behave in a ductile manner if $G/B < 0.5$, otherwise it should be brittle. For ZB-RuN $G/B = 0.63$, thus this phase is expected to behave as a brittle material.

The Young`s modulus ($Y$) and Poisson`s ratio ($\nu$), which belong to the most interesting elastic properties for applications, are also calculated for ZB-RuN using the following relations: $Y=9BG/(3B+G)$ and $\nu =(1-Y/3B)/2$. The data obtained are: $Y$=417.1 GPa and $\nu$ =0.24.

Let us note that the obtained the Poisson`s ratio for ZB-RuN lies in the interval between the $\nu$ values typical for covalent ($\nu \sim 0.1$) and metallic ($\nu \sim 0.33$) materials [15]. This implies that the inter-atomic bonding for ZB-RuN is of mixed covalent-metallic type, see also below. In addition the Cauchy pressure $CP = (C_{12} - C_{44})$ for ZB-RuN is positive ($CP = 112$ GPa) *i.e.* the chemical bonding includes the metallic-like contribution.

Finally, the calculated total densities of states (DOSs) for all RuN polymorphs are depicted in Fig. 2. All of them are characterized with metallic-like electronic spectra having non-zero DOS at Fermi level ($E_F$) and consist of relatively narrow low-lying band (13 – 17 eV below $E_F$) composed mainly with N 2*s* states and wide near-Fermi band (from about -9 eV to $E_F$) which is formed with Ru 4*d* and N 2*p* states which are responsible for Ru-N covalent bonds.

## 4. Conclusion.

In summary, by means of the FLAPW-GGA approach, we have systematically studied the trends in stability, elastic and electronic properties for six possible polymorphs of ruthenium mononitride RuN with various atomic coordination numbers CNs: cubic zinc blende (ZB) and cooperite PtS-like structures with CNs = 4;

cubic rock-salt (RS), hexagonal WC-like and NiAs-like structures with CNs = 6 and cubic CsCl-like structure with CN = 8.

Our main finding is that the most stable is ZB structure, which is much more preferable for RuN than the recently reported RS structure for synthesized RuN samples. The elastic and electronic properties of ZB-RuN were investigated and discussed in comparison with those for RS-RuN polymorph.

**References**

[1] R.B. Kaner, J.J. Gilman, S.H. Tolbert, Science, **308**, 1268 (2005).
[2] Q. Gu, G. Krauss, W. Steurer, Adv. Mater., **20**, 3620 (2008).
[3] A.L. Ivanovskii, Russ. Chem. Rev., **78**, 328 (2009).
[4] M.G. Moreno-Armenta, J. Diaz, A. Martinez-Ruiz, G. Soto. J. Phys. Chem. Solids, **68**, 1989 (2007).
[5] V.V. Bannikov, I.R. Shein, N.I. Medvedeva, A.L. Ivanovskii, J. Magn. Magnet. Mater., **321**, 3624 (2009).
[6] R. de Paiva, R.A. Nogueira, J.L.A. Alves, Rhys. Rev. B, **75**, 085105 (2007)
[7] V.V. Ivanovskaya, I.R. Shein, A.L. Ivanovskii, Diamond Related Mater., **16**, 243 (2007).
[8] E. Zhao, Z. Wu, J. Solid State Chem., **181**, 2814 (2008).
[9] Y. Liang, J. Zhao, B. Zhang. Solid State Commun., **146**, 450 (2008).
[10] P. Blaha, K. Schwarz, G. Madsen et al., *WIEN2k, An Augmented Plane Wave Plus Local Orbitals Program for Calculating Crystal Properties*, Vienna University of Technology, Vienna (2001)
[11] J.P. Perdew, S. Burke, M. Ernzerhof, Phys. Rev. Letters, **77**, 3865 (1996).
[12] P.E. Blochl, O. Jepsen, O.K. Anderson, Phys. Rev. B, **49**, 16223 (1994).
[13] M.J. Mehl , Phys. Rev. B, **47**, 2493 (1993).
[14] S.F. Pugh, Phil. Mag., **45**, 823 (1954).
[15] J. Haines, J.M. Leger, G. Bocquillon, Annu. Rev. Mater. Res. **31**, 1 (2001).

**Table 1.** Calculated lattice constants (*a,c* in Å) and relative total energies (ΔE, in eV/ formula unit) for six RuN polymorphs.

| polymorph | CN* | *a* | *c* | *c/a* | ΔE |
|---|---|---|---|---|---|
| ZB | 4 | 4.555 | - | - | 0 |
| PtS | 4 | 2.692 | 6.042 | 2.244 | 0.465 |
| NiAs | 6 | 2.948 | 5.272 | 1.788 | 0.770 |
| WC | 6 | 2.754 | 3.017 | 1.095 | 1.238 |
| RS | 6 | 4.313 | - | - | 1.276 |
| CsCl | 8 | 2.675 | - | - | 1.762 |

* atomic coordination number

**Table 2.** Calculated elastic constants $C_{ij}$, bulk modules $B$, shear modules $G$ and tetragonal shear modules $G`$ (in GPa) for RuN RS- and ZB-like polymorphs.

| polymorph | $C_{11}$ | $C_{12}$ | $C_{44}$ | $B$ | $G$ | $G`$ |
|---|---|---|---|---|---|---|
| RS | 263.8 | 314.4 | -66.2 | 297.5 | -66.2 | -25.3 |
| ZB | 280.4 | 258.6 | 168.4 | 265.9 | 168.4 | 10.9 |

**Figures**

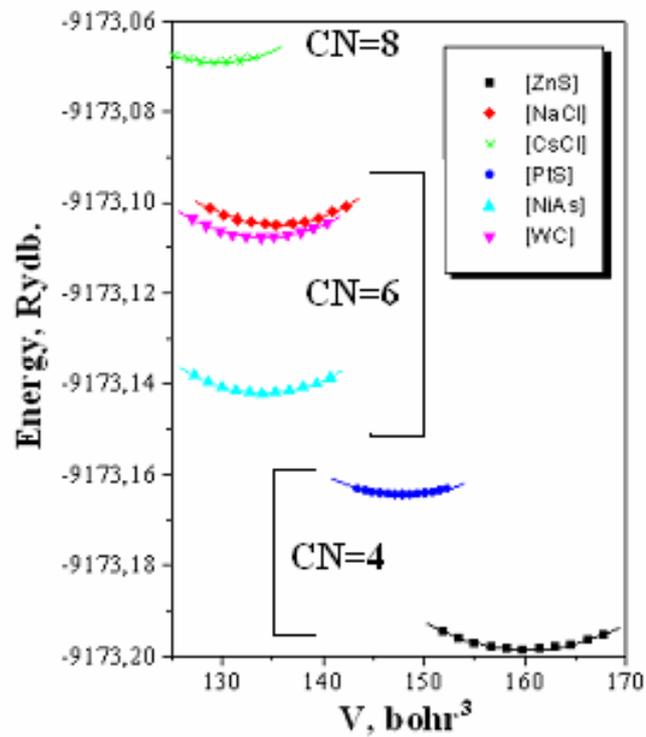

**Figure 1.** The total energy versus cell volume for RuN polymorphs. The atomic coordination numbers (CN) are indicated.

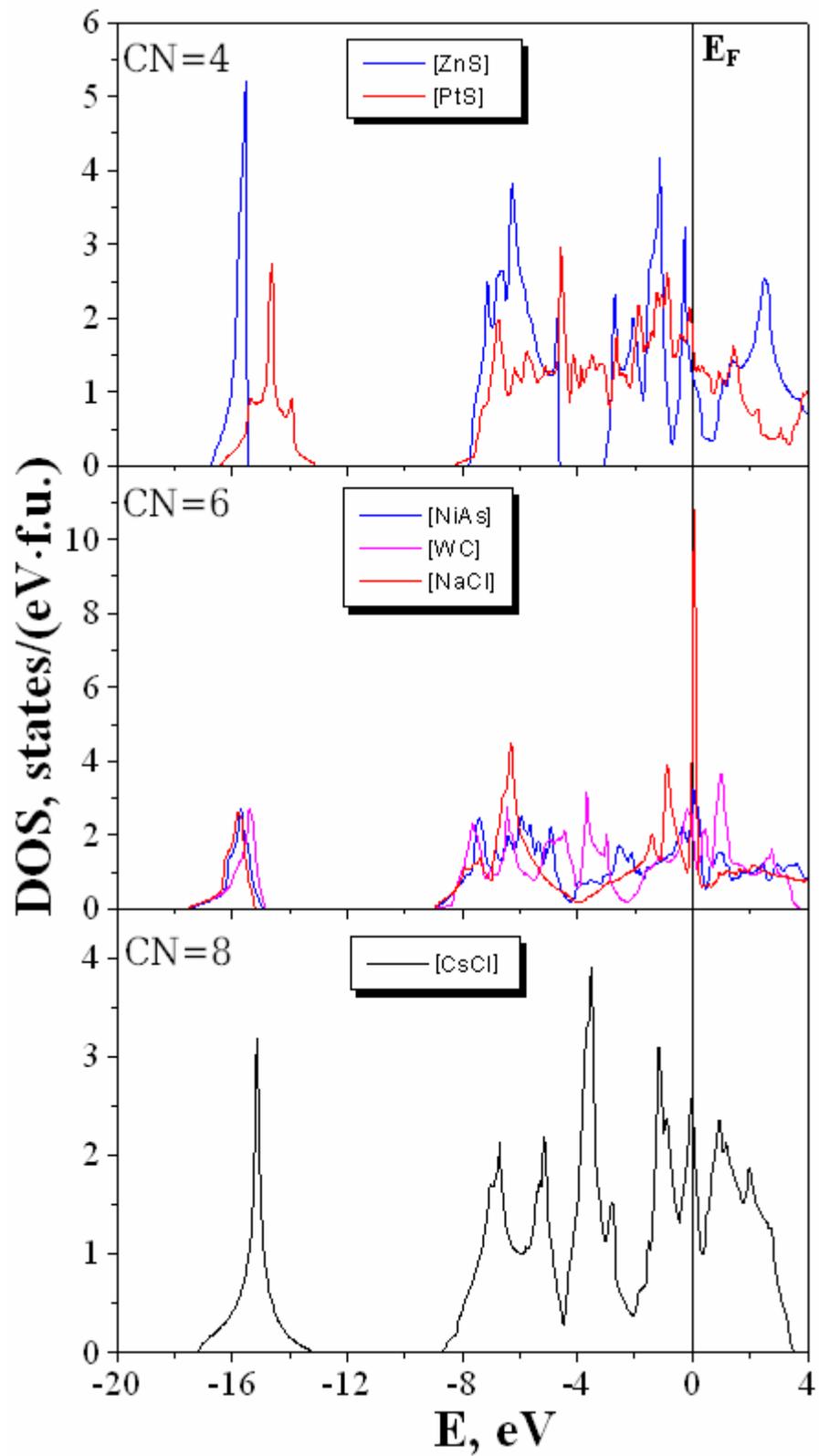

**Figure 2.** Total DOSs for RuN polymorphs grouped by their coordination numbers (CN).

7